\documentclass[twocolumn,showpacs,preprintnumbers,amsmath,amssymb]{revtex4}

\usepackage{graphicx}
\usepackage{dcolumn}
\usepackage{bm}

\begin{document}


\title{Co-transport-induced instability of membrane voltage in tip-growing cells}

\author{M. Leonetti}
\author{P. Marcq}
\author{J. Nuebler}
\affiliation{
IRPHE, Universit\'es Aix-Marseille I and II, UMR CNRS 6594,\\
Technop\^ole de Ch\^ateau-Gombert, 13384 Marseille Cedex 13, France }
\author{F. Homble}
\affiliation{%
Laboratory for Structure and Function of Biological Membranes,
Universit\'e Libre de Bruxelles, CP206/2, Campus Plaine, B-1050
Brussels, Belgium }

\date{September 23, 2005}

\begin{abstract}
A salient feature of stationary patterns in tip-growing cells is the
key role played by the symports and antiports, membrane proteins
that translocate two ionic species at the same time. It is shown
that these co-transporters destabilize generically the membrane
voltage if the two translocated ions diffuse differently and
carry a charge of opposite (same) sign for symports (antiports).
Orders of magnitude obtained for the time and lengthscale
are in agreement with experiments. A weakly nonlinear analysis
characterizes the bifurcation.
\end{abstract}

\maketitle

Spatiotemporal pattern formation of the electric membrane potential
in cells and tissues emerges from collective dynamics and activity
of membrane ion channels. Action potential and cardiac excitation
spiral waves are paradigmatic examples of nonstationary pattern
formation \cite{review,Hohenberg}. Stationary patterns of ionic
currents are widespread in fungi, plant cells (algae for example),
protozoa and insects: Chara corallina, Fucus zygote and Achlya are
the model cells \cite{Gow1989,sitecurrent}. Such patterns are
correlated to cell polarization, apical growth, morphogenesis and
nutrient acquisition. The characteristic wavelengths and times vary
from a few millimeters to ten microns and from one hour to one
minute, respectively. These times correspond typically to a membrane
protein or an ion diffusive time. Two mechanisms have been proposed
\cite{rmq1}: one based on the electromigration of membrane proteins
\cite{FromherzPNAS1988,FromherzPRE1995,LeonettiPRE1997,KreePRE2002,rmq2}
and the other resulting from a negative differential conductance
characterizing voltage-gated channels
\cite{LeonettiPRL1998,LeonettiPNAS2004,rmq}.

\begin{figure}[t]
\includegraphics{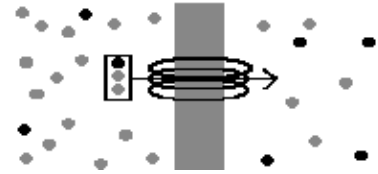}
\caption{\label{fig:1} The pump (not drawn) generates a gradient of
the electrochemical potential through the membrane by translocating
continuously one species of ions (grey disk). This stored free
energy is used by the symport, a co-transporter, to transfer a
second species of ion (black disk) or nutrient against its own
gradient if necessary. The stoechiometry of the drawn symport is
equal to 2: two grey disks for one black disk.}
\end{figure}

However, the origin of current patterns is still unclear in
tip-growing cells where transcellular currents are mainly produced
by the pump and a co-transporter, a membrane carrier that
translocates two species of ions at the same time
\cite{PelcePRL1997,KropfScience1983} (see Fig.~\ref{fig:1}). Three
points of view are proposed by biologists for tip-growing cells:
ionic currents may be a consequence of cellular growth, a
self-organized pattern coupled to growth or, alternatively, arise as
a self-organized pattern which precedes cellular growth
\cite{Harold1990}. The appearance of a lateral branching preceded by
an inward current supports the hypothesis of self-organization in
Achlya \cite{KropfScience1983}. The mechanisms proposed in the
literature cannot explain such patterns \cite{rmq3}.

In this letter, we ask the broader question: how does the stability
of the membrane voltage depend on co-transporters? Only the
contribution of channels to membrane voltage instability has been
investigated in the literature. We demonstrate here that the voltage
along a membrane containing co-transporters is linearly unstable on
a diffusive (not electrical) characteristic time. The final pattern
is a stationary modulation of ionic concentrations, membrane voltage
and transcellular ionic currents. The mechanism is specific to this
kind of carriers since each ionic transporter is characterized by a
\emph{positive} differential conductance.

Consider two ions 1, 2 of valence numbers $z_j$ and concentrations
$C_j$, diffusing along a cylindrical cellular membrane of radius
$r$. As in the cable model, the electrodynamics is governed by a
one-dimensional electrodiffusive equation for each ion:
\begin{equation}
  \label{eq:model1}
\partial_tC_j=D_j\partial^2_xC_j+z_j(eD_j/k_BT)\partial_x(C_j\partial_xV)-(2/r)J_j
\end{equation}
and the capacitive relation for the membrane voltage $V$:
\begin{equation}
  \label{eq:model2}
V=V_0+(Fr/2C_m)(z_1(C_1-C_{10})+z_2(C_2-C_{20}))
\end{equation}
where $D_j$ is the diffusion coefficient of ion j, $V_0$ the resting
membrane potential ($\approx-0.1 \; \mathrm{V}$), $C_m$ the specific
membrane capacitance ($\approx 0.01 \; \mathrm{F m}^{-2}$) and
$C_{j0}$ the concentration of ion j in the resting state. The
standard cable
model is recovered simply from 
(\ref{eq:model1}-\ref{eq:model2}) when all ionic diffusion
coefficients are identical. In our 
model, the fluxes $J_j$ take into account the intracellular chemical
reactions as well as the membrane fluxes through pumps,
co-transporters, channels and uniports. The pump uses chemical
energy (ATP) to translocate ions from one side to the other (the
1-ion in this Letter), generating an electrochemical potential
gradient through the membrane: for example, $\mathrm{H^+}$-ATPase in
plant cells. Consequently, the 1-ion is a cation. Co-transporters
use this stored free molar energy to transfer one 2-ion for $n$
1-ions in the same (opposite) sense for the symport (antiport)
carrier (see Fig.~\ref{fig:1}). In practice, the stoechiometry $n$
is equal to either $1$ or $2$. In tip-growing cells, the 2-ion is
often an essential nutrient for future vegetal metabolism and
consequently, implied in many chemical processes based on enzymatic
binding reactions: metabolism in Achlya hyphae with the methionine
(an amino acid) uptake or the carbon dioxide supply to chloroplasts
for photosynthesis in Chara corallina by HC$\mathrm{O_3^-}$ entry.
Finally, the flux $J_j$ of each ion j is:
\begin{eqnarray}
\label{eq:flux1} J_1 &=& J_{pch} +  n J_s,\\
\label{eq:flux2} J_2 &=& \pm J_s+\alpha C_2+ J_2^{\mathrm{NL}}
\end{eqnarray}
where $J_{pch}$ is the flux through active pumps and passive
channels translocating the 1-ion, $\pm J_s$ ($nJ_s)$ is the flux of
the 2-ion (1-ion) through the co-transporter, and
$J_2^{\mathrm{NL}}$ is the (concentration-dependent) nonlinear part
of $J_2$, due to intracellular chemical processes. The mechanism of
the (linear) voltage instability does not depend on the functional
form of $J_2^{\mathrm{NL}}$. The characteristic kinetic constant
$\alpha$ of nutrient uptake is necessarily positive. The sign $\pm$
is $+$ for a symport and $-$ for an antiport. In the following, we
consider the case of the symport but the extension to that of the
antiport is straightforward. For simplicity's sake, $J_{pch}$ and
$J_s$ do not depend on concentrations and vary linearly with the
membrane potential $V$, characterized by their \emph{positive}
conductances: $G_{pch}=z_1F(\partial J_{pch}/\partial V)$ and
$G_s=z_1F(\partial J_s/\partial V)$. Even if a co-transporter is
often called a secondary active carrier, its working is passive.
Consequently, the differential conductance of the current through
the co-transporter, $(n+z_2/z_1)G_s$ is always positive (positive
Onsager coefficient). Then, there is no local positive feedback
provided by protein characteristics. In the homogeneous resting
state, $J_j=0$ for each ion: the molar flux $J_{s0}$ of the nutrient
uptake in the resting state may be nonzero. Equations
(\ref{eq:model1}-\ref{eq:flux2}) are scaled with dimensionless
coordinates for space $x'=x/\lambda$ and time $t'=t/\tau$ with the
cable length characteristic of the pumps and channels (primes are
then dropped for simplicity), $\lambda^2=r\gamma/2G_{pch}$ and the
diffusive time $\tau=\tilde{D} \lambda^2/D_1D_2$, where $\gamma$ is
the bulk ionic conductivity and $\tilde{D}=\delta_1D_1+\delta_2D_2$
is the mean coefficient of diffusion. We set
$\delta_j=z_j^2C_{j0}/(z_1^2C_{10}+z_2^2C_{20})$ equal to $0.5$ in
all the following.

\begin{figure}
\includegraphics*[width=1.0\columnwidth]{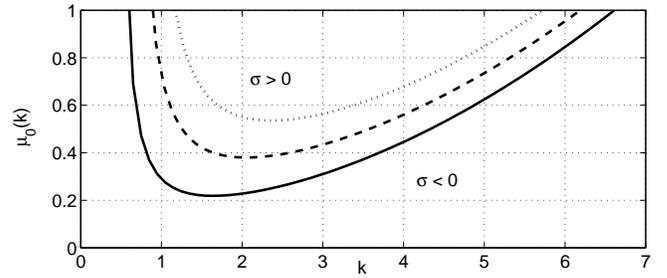}
\caption{\label{fig:neutral} The neutral curve $\mu_0(k)$ is defined
by $\sigma(\mu_0(k);k)=0$ and has a minimum at $(k_c;\mu_c)$. Above
the critical control parameter $\mu_c$, the membrane voltage is
unstable. Parameters are: $n=2$, $z_1=1$, $z_2=-1$,
$D_1=10^{-5}\mathrm{cm}^2\mathrm{s}^{-1}$,
$D_2=10^{-7}\mathrm{cm}^2\mathrm{s}^{-1}$ (common for the three
neutral curves), and $\beta_1=0.1$(-), $\beta_1=0.2$(- -) and
$\beta_1=0.3$(..). }
\end{figure}

The control parameter $\mu$ is the positive conductance ratio,
$\mu=G_s/G_{pch}$ that controls the ionic membrane fluxes. The
stability of the homogeneous equilibrium state is analyzed by
considering the evolution of fluctuations of voltage $V$ and ionic
concentrations $C_j$ and consequently, the linearized equations of
(\ref{eq:model1}-\ref{eq:flux2}): $\delta H(x,t)=\delta H_0 e^{st +
ikx}$ where $k$ is the wave number of the perturbation and $H$
refers to $V$ and $C_j$. Two real solutions for $s = s(k)$ are
determined: $Im(s)=0$, $Re(s)=\sigma(\mu;k)$. The first one is the
well-known fast capacitive relaxation. The second one yields the
growth rate of the instability:
\begin{eqnarray}
\sigma(\mu;k)&=&-\frac{k^4+\frac{\tilde{D}}{D_1}k^2\Big(1+\mu\Big(n+\frac{D_1z_2}{D_2z_1}\Big)\Big)}
{k^2+1+(n+z_2/z_1)\mu}\label{eq:dispersion}\\
&&-\frac{\beta_1}{D_2}\frac{(\tilde{D}-D_2)k^2+(D_1-D_2)(1+n\mu)\frac{\tilde{D}}{D_1}}{k^2+1+(n+z_2/z_1)\mu}\nonumber
\end{eqnarray}
where $\beta_1=\alpha\gamma/G_{pch}(D_1-D_2)$ is dimensionless.
Since the capacitance does not appear in (\ref{eq:dispersion}),
this characteristic (inverse) time is \emph{diffusive}. From
$\sigma(\mu_0(k);k)=0$, the neutral curve $\mu_0(k)$ is determined
and has a minimum defining the critical values of the control
parameter $\mu_c$ and the wavenumber $k_c$ (Fig.~\ref{fig:neutral}):
\begin{equation}
\nonumber
\mu_c=\frac{-D_1D_2}{\tilde{D}(z_2D_1/z_1+nD_2)}\left(2k_c^2+\frac{\tilde{D}}{D_1}+
\beta_1(\frac{\tilde{D}}{D_2}-1)\right)
\end{equation}
\begin{equation}
\nonumber
(k_c^2-k_0^2)^2=k_0^4+\frac{\beta_1 \tilde{D}}{D_1 D_2}  (D_1-D_2)+
k_0^2\left(\frac{\tilde{D}}{D_1}+\beta_1(\frac{\tilde{D}}{D_2}-1)\right)
\end{equation}
where $k_0^2=-n\beta_1(D_1-D_2)/(z_2D_1/z_1+nD_2)$. The particular
case $\alpha = \beta_1 = 0$ corresponds to a long-wavelength
instability ($k_c = 0$), and will not be treated here \cite{KreePRE2002}.

For $\mu > \mu_c > 0$, the growth rate is positive in a finite range
of wavenumbers. The limit case of small binding reactions may help
clarify the nature of this instability. For small but nonzero
$\beta_1$, the homogeneous resting state
is unstable ($\sigma > 0$) against 
spatial perturbations if $1+\mu(n+D_1z_2/D_2z_1)<0$. Recalling that
$0\leq n+z_2/z_1$, two necessary conditions for instability are $D_1
> n D_2 |\frac{z_1}{z_2}|$ and $z_2/z_1 < 0$  \cite{rmq4}. The
former is generally verified since binding reactions reduce notably
the effective diffusion of the 2-ion \cite{rmq5}; the latter is
fulfilled for the symports $\mathrm{H}^+/\mathrm{HCO}_3^-$ in Chara
corallina and $\mathrm{H}^+$-methionine in Achlya. For large control
parameter $\mu$ and small $\beta_1$, the wavelength $\lambda_p$ of
the pattern satisfies: $\lambda_p \approx 2\pi (r\gamma D_2 / 2G_s
\tilde{D})^{1/2} \approx \lambda_{\mathrm{cable}}
(D_2/\tilde{D})^{1/2}$ where $\lambda_{\mathrm{cable}}$ is provided
by the cable model and may be experimentally measured by two impaled
electrodes. An order of magnitude of $\lambda_p$ can thus be
evaluated. In Achlya hyphae, some measurements indicate
$\lambda_{\mathrm{cable}} \approx 2 \; \mathrm{mm}$
\cite{Kropf1986}. On the basis of measurements of the effective
diffusion of the calcium ion, a reasonable value for the diffusion
coefficient of the 2-ion is $D_2 \approx 10^{-7}
\mathrm{cm}^2/\mathrm{s}$. We thus expect a characteristic pattern
wavelength $\lambda_p \approx 100 \mu \mathrm{m}$, in agreement with
experiments. The characteristic time $\tau_p$ required to produce
the pattern is of the order of a diffusive time: $\tau_p \approx
\lambda_p^2/D_2$, dominated by the slower ionic diffusion. Using the
previous values, we find $\tau_p \approx 10^3 \mathrm{s}$,  in
agreement with experiments on Achlya hyphae \cite{Gow1989}.

\begin{figure}
\hfill
\includegraphics*[width=1.0\columnwidth]{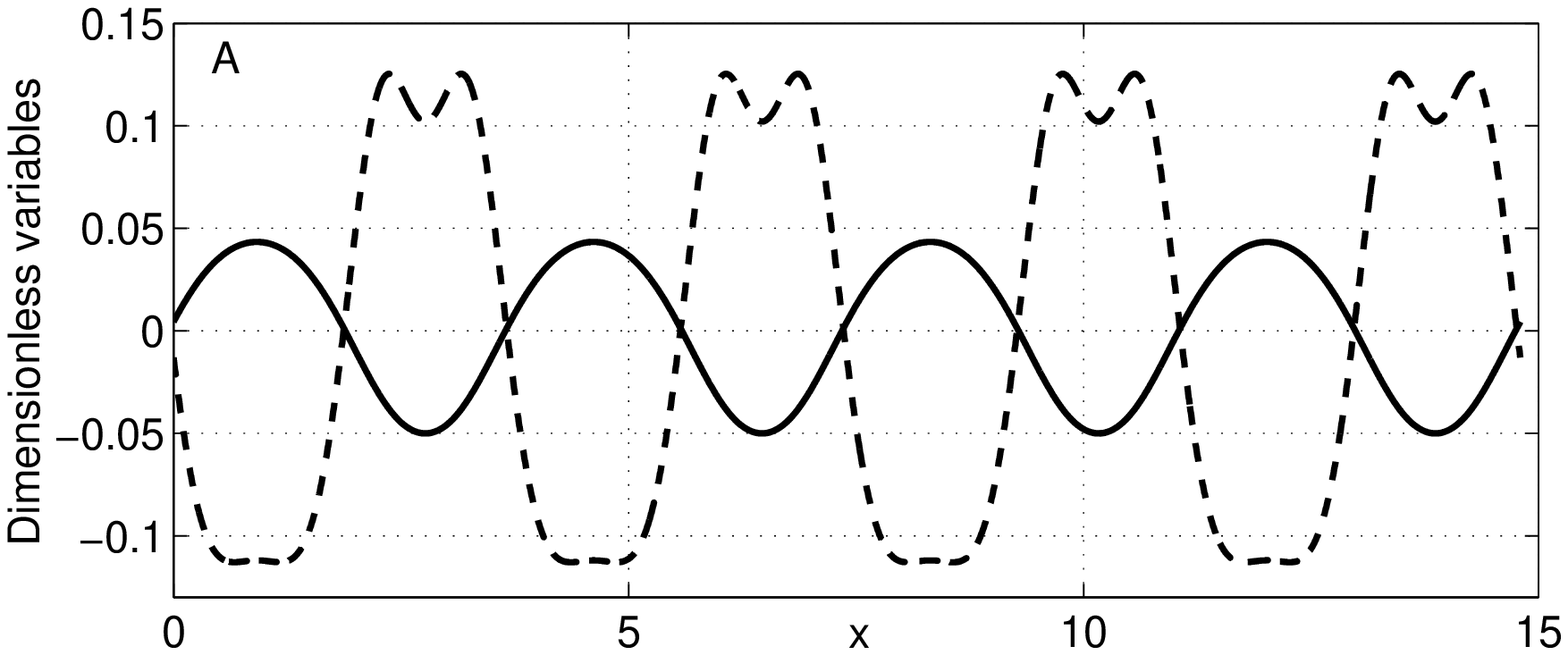}
\hfill
\includegraphics*[width=1.0\columnwidth]{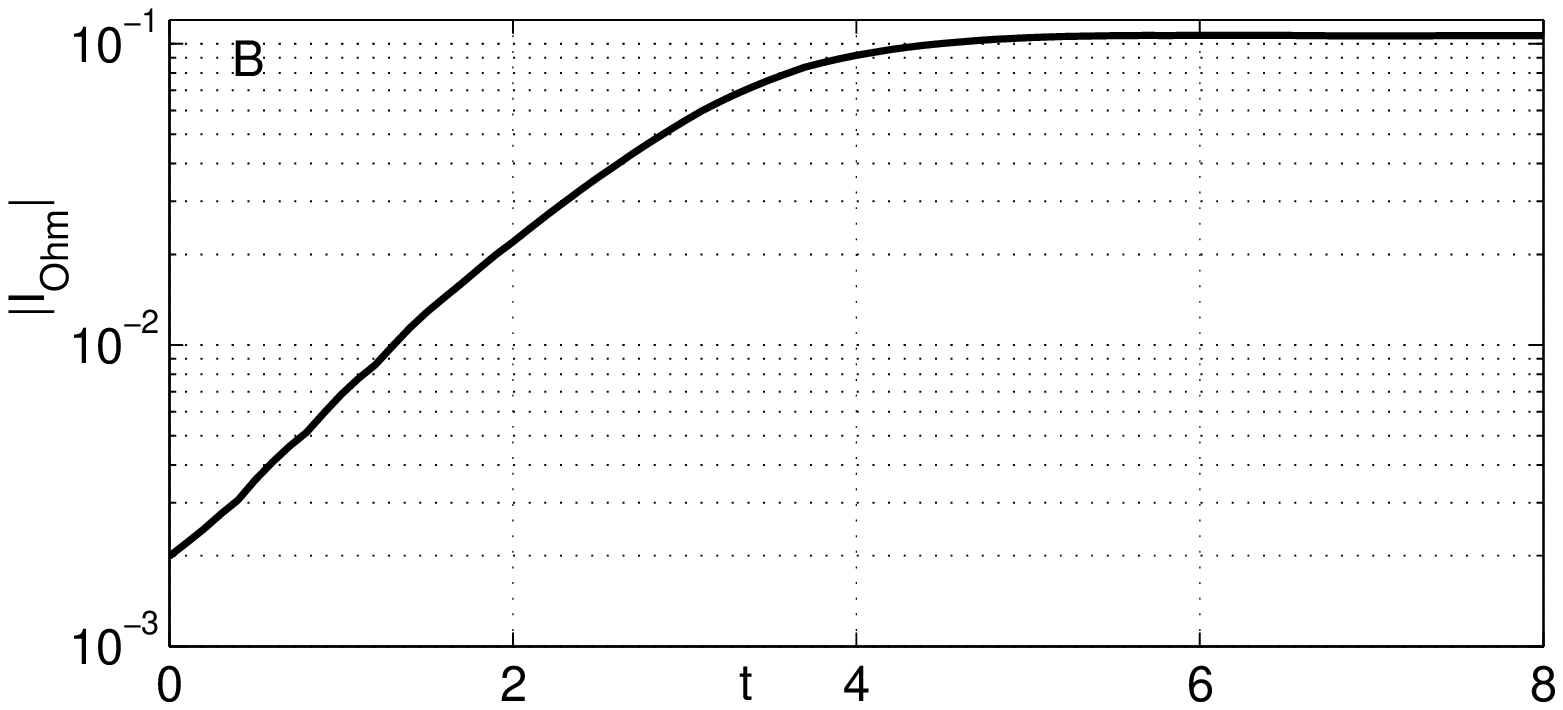}
\hfill
\caption{\label{fig:pattern}
Parameters are the same as in
Fig.~\ref{fig:neutral}, with in addition $\mu=0.25$, $\beta_1=0.1$,
$\beta_2=0.1$ and $\beta_3=5$. \textbf{A}: The final stationary
pattern is a modulation of the dimensionless membrane potential
$(V-V_0)/|V_0|$ (-) and of the dimensionless ohmic part
$I_{\mathrm{Ohm}}$ (- -) of the extracellular current. An
hyperpolarized membrane potential $(V-V_0)/|V_0|<0$ corresponds to
an outer ohmic current (electric field) in agreement with
experiments made with the vibrating probe. \textbf{B}: Temporal
evolution of the extracellular current $I_{\mathrm{Ohm}}$ at the
position $x=12.5$. The characteristic time is an ionic diffusive
one.}
\end{figure}

Equations (\ref{eq:model1}-\ref{eq:flux2})
have been solved numerically for a large range of parameters to 
confirm the previous results.
For simplicity, the nonlinear flux $J_2^{\mathrm{NL}}$ is given
by a truncated expansion in powers of the concentration of the 2-ion:
\begin{eqnarray}
  \label{eq:j2nl}
J_2^{\mathrm{NL}} = C_{20} \sum_{j=2,3}\alpha_j
((C_2-C_{20})/C_{20})^j
\end{eqnarray}
This form generalizes the expansion of a Michaelis-Menten enzyme
kinetics term in the limit of large Michaelis constant: our goal is
to take into account at a phenomenological level some of the
complexity due to the function of the 2-ion. The coefficient
$\alpha_3$ must be positive to ensure nonlinear convergence. The
simulation depends on two additional dimensionless parameters:
$\beta_2=\gamma^2 |V_0|\alpha_2/(z_2FG_{pch}C_{20}(D_1-D_2)^2)$ and
$\beta_3=\gamma^3
|V_0|^2\alpha_3/((z_2F)^2G_{pch}C_{20}^2(D_1-D_2)^3)$. Generally,
the voltage relaxes to zero on a characteristic capacitive time, as
expected from the cable model. However, for relevant parameters
characterizing a symport, a cellular pattern of voltage and
concentrations appears after a transient whose duration is of the
order of magnitude of the diffusive time $\tau_p$
(Fig.~\ref{fig:pattern}). Outer and inner transcellular currents
flow periodically through the membrane. It has been established that
the ohmic part $I_{\mathrm{Ohm}}$ of the dimensionless extracellular
current
 normal to the membrane is given by the relation:
$I_{\mathrm{Ohm}}=\tilde{D}(I_1/D_1+I_2/D_2)/G_{pch}|V_0|$
\cite{LeonettiPNAS2004}. An outer (inner) ohmic current corresponds
to an hyperpolarized (depolarized) band in agreement with
experiments (Fig.~\ref{fig:pattern}). The outer current has a
characteristic M-shape, observed in Chara corallina. Varying the
nonlinear parameters, it is possible to obtain a M-shape only for
the inner current or for both.

\begin{figure}
\includegraphics*[width=1.0\columnwidth]{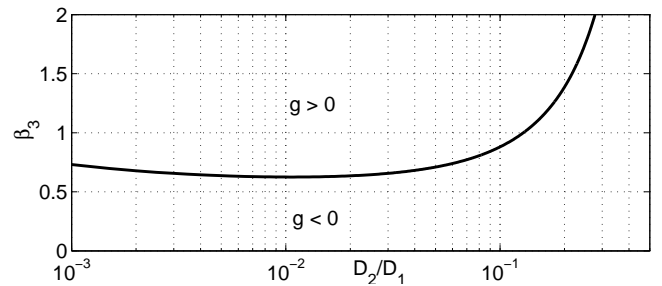}
\caption{\label{fig:bif} The tricritical line $g=0$ separates
domains in the reduced parameter space $(D_2/D_1;\beta_3)$ where the
bifurcation is supercritical ($g>0$) and subcritical ($g<0$). The
fixed parameter values are the same as in Fig.~\ref{fig:pattern}.}
\end{figure}

The stationary bifurcation is further characterized by a weakly
nonlinear analysis performed in the vicinity of the threshold $(k_c;
\mu_c)$ \cite{Hohenberg}. An arbitrarily small expansion parameter
$\epsilon$ is introduced to separate the fast and slow scales in the
problem. We define the slow independent variables $X=\epsilon x$ and
$T=\epsilon^2 t$, and Taylor-expand the concentrations $C_j$,
membrane voltage $V$ and control parameter $\mu$ in powers of
$\epsilon$. The resulting equations are then solved recursively for
each power $\epsilon^i$. The solvability condition (Fredholm
alternative) at third order provides the amplitude equation:
\begin{equation}
  \label{eq:amplitude}
\tau_0 \partial_T A= \bar{\mu} A+\xi^2_0\partial^2_X A-g |A|^2A,
\end{equation}
where $\bar{\mu} = (\mu-\mu_c)/\mu_c$ is the reduced control
parameter. The time and lengthscale $\tau_0$ and
$\xi_0$ of the pattern's slow modulations close to the bifurcation
may also be derived directly fom the dispersion relation
(\ref{eq:dispersion}): $\mu_c \xi^2_0 =
\frac{1}{2} \; \left(\frac{\partial^2 \mu_0}{\partial k^2}\right)_c$
and $\tau_0^{-1}= \mu_c \left(\frac{\partial \sigma}{\partial
\mu}\right)_c$ \cite{Hohenberg}. The coefficient $g$ of the nonlinear term is a
complicated function of the physical parameters. The bifurcation is
supercritical (resp. subcritical) for positive (resp. negative)
values of $g$. In the idealized case described here, and for typical
parameter values, a tricritical line $g = 0$ separates the two types
of bifurcation in parameter space (see Fig.~\ref{fig:bif}).

In conclusion, we established that a spatially homogeneous membrane
voltage is linearly unstable if the co-transporters play a role in
the control of the electrophysiological properties of a cell. The
final stationary pattern is a transcellular current bearing various
ionic species. As opposed to many other scenarios leading to
spatiotemporal pattern formation in ionic currents
\cite{review,LeonettiPRL1998,LeonettiPNAS2004}, a negative
differential conductance is not required. A necessary ingredient is
the slow intracellular diffusion of one of the two ions translocated
by the co-transporter. Interestingly, this is often the case in
experiments. This mechanism may explain how a cell can uptake an
essential nutrient at precise locations: in Achlya, methionine
enters at the tip during apical growth.

We thank P. Pelc\'e for useful discussions.
We wish to acknowledge the support of ``ACI Physico-chimie des
syst\`emes complexes'' (France), the Fonds National de la
Recherche Scientifique (Belgium) and the Communaut\'e Fran\c{c}aise de
Belgique-Action de Recherches Concert\'ees (Belgium).


\begin{thebibliography}{article}
\bibitem{rmq1}
Pattern formation of transcellular ionic currents could also result
from a Turing instability. However, contrary to the two other
mechanisms, experimental evidence is still lacking.
\bibitem{rmq2}
Experiments on the effect of electric fields on clustering of
Acetylcholine receptors proves the relevance of the Fromherz
mechanism at scales less than 50 microns. However, the large
characteristic time predicted for patterns of larger wavelengths,
such as observed in Chara corallina, is not consistent with
experiments.
\bibitem{rmq4}
For antiports, $n+D_1z_2/D_2z_1<0$ is replaced by \mbox{$n-D_1z_2/D_2z_1<0$}.
The two necessary conditions are $D_1 > n D_2 |\frac{z_1}{z_2}|$
(unchanged) and $z_2/z_1>0$ (often fulfilled by antiports).
\bibitem{rmq5}
For example, it is well established that the diffusion constant of
the calcium ion is then reduced by at least a factor $100$.
\end{thebibliography}
\end{document}